\documentstyle[twoside,fleqn,espcrc2,epsfig]{article}

\newcommand{\beq}{\begin{equation}}
\newcommand{\eeq}{\end{equation}}
\newcommand{\beqn}{\begin{eqnarray}}
\newcommand{\eeqn}{\end{eqnarray}}
\newcommand{\bea}[1]{\beq\begin{array}{#1}}
\newcommand{\eea}{\end{array}\eeq}
\newcommand{\summ}[2]{\sum\limits_{#1}^{#2}}

\newcommand{\Z}{{Z \!\!\! Z}}

\newcommand{\dual}{\mbox{}^{\ast}}
\newcommand{\diff}{\partial}
\newcommand{\concorr}[1]{<\!\!\!<#1>\!\!\!>}
\newcommand{\corr}[1]{<\!\!#1\!\!>}
\def\NP{ Nucl.~Phys.}
\def\PL{ Phys. Lett.}
\def\PR{ Phys.~Rev.}

\title{
\vspace{-3.6cm}
\begin{flushright}
{\normalsize
ITEP-TH-44/97\\
\vspace{-.2cm}
September 97}
\end{flushright}
\vspace{1.5cm}
Electric and magnetic currents in
$SU(2)$ lattice gauge theory\thanks{Talk given by F.V.~Gubarev at 
the International Symposium on Lattice Field Theory, 22-26 July 
1997, Edinburgh, Scotland}} 
\author{ M.N.~Chernodub$^a$, 
F.V.~Gubarev$^a$, M.I.~Polikarpov\address{ ITEP, B.Cheremushkinskaya 
25, Moscow, 117259, Russia}}

\begin{document}

\begin{abstract}
The correlations of the topological charge~($Q$), the
electric~($J^e$) and the magnetic~($J^m$) currents in the $SU(2)$
lattice gauge theory in the Maximal Abelian projection are
investigated. A nonzero value of the correlator $<QJ^eJ^m>$ is
obtained for a wide range of values of the bare charge, as well as
under the cooling.
\end{abstract}

\maketitle

\section{INTRODUCTION}

An oldest and rather popular model of the QCD vacuum is the
instanton--anti-instan\-ton media
(see~\cite{InstantonicReviews} and the  references therein).
It is not clear  however, whether it possible to explain
the confinement phenomenon within this
approach~\cite{AHazenfratz}, \cite{Diakonovetall}.

On the other hand,  the method of abelian projections~\cite{t'Hooft}
is widely used in numerical calculations to study the confinement
mechanism. It is clearly seen~\cite{MIP_on_latt96}
that the vacuum of the lattice
gluodynamics behaves like  a dual superconductor in the so-called
Maximal Abelian (MaA) projection~\cite{KrScWi87,AboutMaag}.
There are clear indications that abelian monopoles are condensed
in the confinement phase of lattice
gluodynamics~\cite{MonopoleCondensation}.

It occurs that instanton-like configurations and monopoles in the MaA
gauge are interrelated~\cite{FirstAboutIM}.  The relation between
monopoles and instantons has been established
analytically in~\cite{Classical1}, \cite{Classical2} and
numerically in~\cite{Austria}.

In the field of a  single instanton
the monopole currents in the MaA  projection are accompanied
by electric currents~\cite{BornSchierholz}. The qualitative
explanation of this fact is simple. Consider the (anti)self-dual
configuration
\beq                 \label{duality}
F_{\mu\nu}(A)=\pm\dual
F_{\mu\nu}(A)\;.
\eeq
The MaA projection is defined~\cite{KrScWi87} by
the minimization  of the functional $R[A^{\Omega}(x)]$ over the gauge
transformations $\Omega(x)$,
$R[A]=\int d^4x[(A_{\mu}^{1})^2+(A_{\mu}^{2})^2]$,
so that in the MaA gauge one can expect
the  abelian component of the commutator term
$1/2Tr(\sigma^3[A_{\mu},A_{\nu}])=\varepsilon^{3ab}A^a_{\mu}A^b_{\nu}$
to be  small compared with the abelian field-strength
$f_{\mu\nu}(A)=\diff_{[\mu}A^3_{\nu ]}$.
Therefore, in the MaA projection eq.(\ref{duality}) yields
\beq                                          \label{abelian_duality}
f_{\mu\nu}(A)\approx\pm\dual f_{\mu\nu}(A)\,.
\eeq
Due to eq.(\ref{abelian_duality}),  the monopole currents have to be
correlated with the electric ones since
\beq                                        \label{jm_je}
J^e_{\mu}=\diff_{\nu} f_{\mu\nu}(A)\approx
\pm\diff_{\nu}\dual f_{\mu\nu}(A)=J^m_{\mu}\;.
\eeq

In the present publication  we study the correlation of electric and
magnetic currents in the real vacuum of the $SU(2)$ lattice
gluodynamics.

\section{MAGNETIC AND ELECTRIC CURRENTS IN THE
ABELIAN PROJECTION OF THE $SU(2)$ LATTICE GLUODYNAMICS}

The abelian monopoles exist
in the abelian projection,  since the residual $U(1)$ group is compact.
The definition of the abelian monopole current is~\cite{DGT}:
\beq                                     \label{monopole_current}
J^m_{\mu}(y)=\frac{1}{4\pi} \sum\limits_{\nu\lambda\rho}
\varepsilon_{\mu\nu\lambda\rho}
[\bar\theta_{\lambda\rho}(x+\hat{\mu})-\bar\theta_{\lambda\rho}(x)].
\eeq
Here the angle $\bar\theta_{\mu\nu}$ is the normalized plaquette angle
$\theta_{\mu\nu}=\bar\theta_{\mu\nu}+2\pi k_{\mu\nu}$;
$k_{\mu\nu}$ is an integer such that
$\bar\theta_{\mu\nu}\in (-\pi;\pi]$.
The monopole currents are quantized ($J^m_{\mu}\in \Z$) and
conserved ($\diff_{\mu} J^m_{\mu}=0$).
They are attached to the links of the dual lattice,
the link $(y,\mu)$ is dual to the cube $(x,\nu\lambda\rho)$.

The electric current is defined as
\beq                                      \label{electric_current}
J^e_{\mu}(x)=\frac{1}{2\pi} \sum\limits_{\nu}
[\bar\theta_{\mu\nu}(x)-\bar\theta_{\mu\nu}(x-\hat{\nu})].
\eeq
In the continuum limit,  the definitions~(\ref{monopole_current})
and~(\ref{electric_current})  correspond to the usual ones:
$J^m_{\mu}=\diff_{\nu}\dual f_{\mu\nu}$;
$J^e_{\mu}=\diff_{\nu} f_{\mu\nu}$.
The electric currents are conserved
($\diff_{\mu} J^e_{\mu}=0$) and attached to the links  of the
original lattice.  Electric currents are not
quantized.

In order to calculate the correlators of the type
$\corr{J^e_{\mu}(x)J^m_{\mu}(x)...}$
one has to define the electric current on the dual lattice
or the  magnetic current on the original lattice.
We define the electric current on the dual lattice in
the following way:
\beq                                         \label{correspondence}
J^e_{\mu}(y) = \frac{1}{16}\summ{x\in \dual C(y,\mu)}{} \left[
J^e_{\mu}(x)+J^e_{\mu}(x-\hat{\mu})\right]\;.
\eeq
Here,  the  summation in
r.h.s. is over eight vertices $x$ of the 3-dimensional cube $\dual C(y,\mu)$,
to which the current $J^e_{\mu}(y)$ is dual. The point $y$ lies on the dual
lattice and the points $x$ lie on the original one.

For the topological charge density operator
we use the simplest definition:
\beqn
Q(x) & = & \frac{1}{2^9 \pi^2} \sum_{\mu_1,...,\mu_4=-4}^{4}
\nonumber\\
& & \varepsilon^{\mu_1,...,\mu_4}
Tr[U_{\mu_1,\mu_2}(x) U_{\mu_3,\mu_4}(x)],
\eeqn
where $U_{\mu_1,\mu_2}$ is the plaquette matrix. On the dual lattice
the topological charge density corresponding to the monopole current
$J^m_{\mu}(y)$ is defined by taking the average over the eight sites
nearest to the current $J^m_{\mu}(y)$:
\beq
Q(y)=\frac{1}{8}\summ{x}{} Q(x)
\eeq

The simplest (connected) correlator of electric and magnetic currents is
\beqn
\concorr{J^m_{\mu}J^e_{\mu}} & = &
\corr{J^m_{\mu}J^e_{\mu}} \nonumber\\
& - & \corr{J^m_{\mu}}\corr{J^e_{\mu}}
\equiv \corr{J^m_{\mu}J^e_{\mu}}\;.
\eeqn
This correlator is equal to zero, since $J^m_{\mu}$ and $J^e_{\mu}$
have  opposite parities.
A  scalar quantity  can be constructed if we multiply $J^m_{\mu}J^e_{\mu}$
by the density of topological charge. The corresponding irreducible
correlator
\beq\label{JmJeQ}
\concorr{J^m_{\mu}J^e_{\mu}Q} \equiv \corr{J^m_{\mu}J^e_{\mu}Q}
\eeq
is nonzero for the vacuum consisting of (anti-)self-dual
domains~({\it cf.} eq.~(\ref{jm_je})).

\section{NUMERICAL RESULTS}

The $SU(2)$ lattice gauge theory was considered on the  $8^4$ lattice with
the Wilson action. At each value of $\beta$ we thermalize lattice fields
using the standard heat bath algorithm.  We calculate the correlators
$\corr{J^m_{\mu}(y) J^e_{\mu}(y)}$,
$\corr{J^m_{\mu}(y) J^e_{\mu}(y) Q(y)}$,
using 100 statistically
independent configurations at each value of $\beta$.

These correlators strongly depend on $\beta$ and it is convenient to
normalize them,  dividing by $\rho^m \rho^e$. Here
$\rho^m$ and $\rho^e$ are the monopole and the electric current  densities:
\beq
\rho_{m(e)}=\frac{1}{4V}\summ{l}{} |J^{m(e)}_l|,
\eeq
$V$ is the lattice volume (the total number of sites).

\begin{center}
\begin{figure}[htb]
\epsfig{file=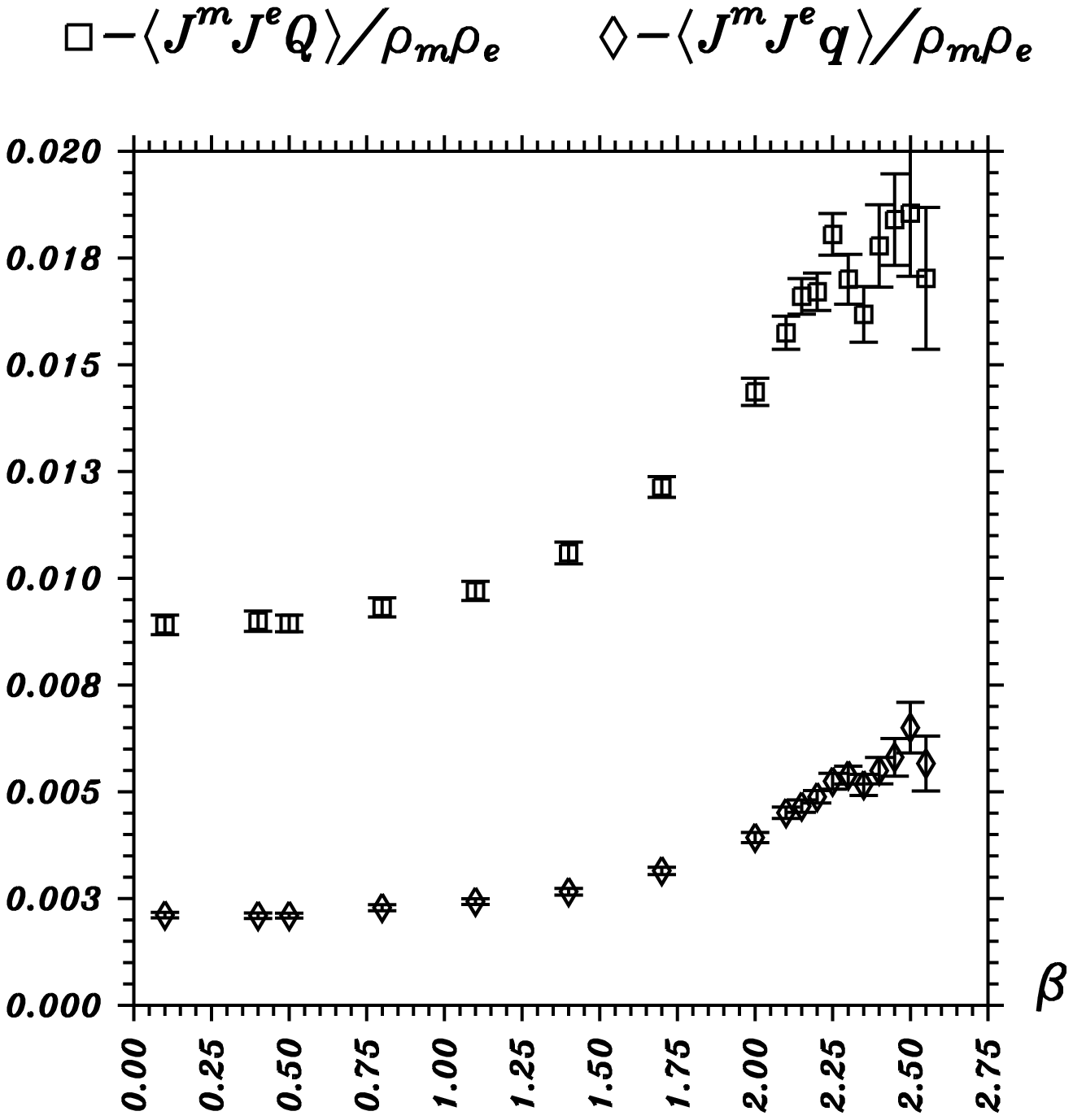,height=7.0cm}
\vspace{-.3cm}
\caption{
Correlators $<\!\!J^m J^eQ\!\!>\!\!/\rho_m\rho_e$
and $<\!\!J^mJ^eq\!\!>\!\!/\rho_m\rho_e$
as a function of $\beta$.
}
\vspace{-.8cm}
\end{figure}
\end{center}

The correlators $\corr{J^mJ^eQ}\!\!/\rho_m\rho_e$ and
$\corr{J^mJ^eq}\!\!/\rho_m\rho_e \quad (q(y)=Q(y)/|Q(y)|)$
are represented in  Fig.1. As one can see from Fig.~1,  the product
of the electric and the magnetic currents is correlated with the topological
charge. Numerical simulations show that  this correlation
increases  under the cooling. The last fact is in agreement with the
results of \cite{BornSchierholz}: the cooled vacuum is populated
by instantons which induce the electric charge to the abelian
monopoles.

\section{CONCLUSION~AND~ACKNOWLEDG\-MENTS}

Our results show that in the vacuum of lattice  gluodynamics the
magnetic current is correlated with the electric current.  Thus the
abelian monopoles have  electric charge. The sign of the electric
charge depends on the sign of the topological charge density.

M.N.Ch and M.I.P. acknowledge the kind hospitality of the Theoretical
Department of the  Kanazawa University.  F.V.G. is grateful for the
kind hospitality of the Theoretical Physics Department of the Vrije
University of Amsterdam. This work has been supported by the JSPS
Program on Japan -- FSU scientists collaboration, and also by the
Grants:  INTAS-94-0840, INTAS-94-2851, INTAS-RFBR-95-0681,  and Grant
No.  96-02-17230a  of  the Russian Foundation for Fundamental
Sciences.


\begin{thebibliography}{50}

\bibitem{InstantonicReviews}
T.Schaefer, E.V.Shuryak, {\it 'Instantons in QCD'},
{\tt hep-ph/9610451}, to appear in {\it Rev. Mod. Phys.};\\
T.Schaefer, E.V.Shuryak, {\it Phys. Rev.}{\bf D53} (1996) 6522.


\bibitem{Diakonovetall}
D.I.Diakonov, V.Yu.Petrov, {\it 'Confinement From Instantons?'}.
Talk given at International
Workshop on Nonperturbative Approaches to QCD, Trento, Italy, 10-29 Jul
1995.

\bibitem{MIP_on_latt96}
M.I. Polikarpov, {\it Nucl.Phys.(Proc.Suppl.)} {\bf 53} (1997) 134,
{\tt hep-lat/9609020}


\bibitem{AHazenfratz}
T.DeGrand, A.Hasenfratz, T.G. Kovacs,
{\it 'Topological Structure in the SU(2) Vacuum'},
{\it preprint COLO-HEP-383}, {\tt hep-lat/9705009}.


\bibitem{t'Hooft}
G.~'t~Hooft, {\it \NP}{\bf B190[FS3]} (1981) 455.


\bibitem{FirstAboutIM}
O.Miyamura, S.Origuchi, {\it 'QCD monopoles and Chiral Symmetry
Breaking in $SU(2)$ Lattice Gauge Theory'},
RCNP Confinement 1995, Osaka, Japan, Mar 22-26, 1995,
p.137.


\bibitem{Classical1}
M.N.Chernodub, F.V.Gubarev, {\it JETP Lett.} {\bf 62} (1995) 100; \\
R.C.Brower, K.N.Orginos, Chung-I Tan, {\it Phys. Rev.}{\bf D55}
(1997) 6313.

\bibitem{Classical2}
A.Hart, M.Teper, {\it \PL} {\bf 371} (1996) 261;  \\
M.Fukushima {\it et al.},
{\it Phys. Lett.} {\bf B399} (1997) 141.

\bibitem{BornSchierholz}
V.Bornyakov, G.Schierholz, {\it \PL} {\bf 384} (1996) 190;


\bibitem{Austria}
S.Thurner {\it et al.},
{\it \PR} {\bf D54} (1996) 3457;   \\
M.Feurstein, H.Markum, S.Thurner, {\it \PL} {\bf B396} (1997) 203-209.


\bibitem{KrScWi87}
A.S~Kronfeld, G.Schierholz and U.-J.~Wiese, {\it \NP}, {\bf B293} (1987)
461. \\
U.J.~Wiese, {\it Phys.Lett.} {\bf 198B} (1987) 516.


\bibitem{AboutMaag}
M.N. Chernodub, M.I. Polikarpov, A.I. Ve\-selov,
{\it Phys. Lett.}{\bf B342} (1995) 303;\\
S.Ejiri {\it et al.}, {\tt hep-lat/9509013},
{\it Lattice~1995}, p.322.

\bibitem{MonopoleCondensation}
T.L. Ivanenko, A.V. Pochinskii and M.I. Polikarpov,
{\it Phys.Lett.} {\bf B302} (1993) 458;\\
L.~Del~Debbio, A.~Di~Giacomo, G.~Paffuti and P.~Pieri,
\PL {\bf B355} (1995) 255;\\
N.~Nakamura {\it et al.}, {\it
Nucl.Phys. (Proc.Suppl.)} {\bf 53} (1997) 512,
{\tt hep-lat/9608004};\\
M.N.~Chernodub, M.I.~Polikarpov and A.I. Veselov,
\PL {\bf B399} (1997) 267.

\bibitem{DGT}
T.A.~DeGrand and D.~Toussaint, {\it Phys.Rev.} {\bf D22} (1980) 2478.

\end{thebibliography}
\end{document}